\documentclass[useAMS,usenatbib]{mn2e}
\usepackage{rotating}
\usepackage{lscape}
\title[Chemical composition of evolved stars in IC~4651]
      {Chemical composition of evolved stars in the open cluster IC~4651\thanks{Based on observations collected at ESO telescopes under 
programmes 65.N-0286 and in part 169.D-0473}}

\author[\v{S}ar\={u}nas Mikolaitis et al.]
       {\v{S}ar\={u}nas Mikolaitis,$^{1}$\thanks{E-mail: sarunas.mikolaitis@tfai.vu.lt}
       Gra\v{z}ina Tautvai\v sien\. e,$^{1}$ 
       Raffaele Gratton,$^{2}$
       Angela Bragaglia$^{3}$ 
       \newauthor and Eugenio Carretta$^{3}$\\ 
$^{1}$Institute of Theoretical Physics and Astronomy, Vilnius University, Go\v{s}tauto 
12, Vilnius 01108, Lithuania\\
       $^{2}$INAF - Osservatorio Astronomico di Padova, Vicolo dell'Osservatorio 5, I-35122 Padova, Italy\\
       $^{3}$INAF - Osservatorio Astronomico di Bologna, Via Ranzani 1, I-40127 Bologna, Italy}
       
\begin{document}

\date{Accepted 2010 .....; Received 2010 .....; in original form 2010 ......}

\pagerange{\pageref{firstpage}--\pageref{lastpage}} \pubyear{2010}

\maketitle

\label{firstpage}

\begin{abstract}
We present an analysis of high-resolution spectra of three core-helium-burning 
`clump' stars and two first ascent giants in the open cluster IC~4651. 
Atmospheric parameters ($T_{\rm eff}$,  log~$g$,  $ v_{\rm t}$, and [Fe/H]) were
determined in our previous  study by Carretta et al.\ (2004). In this study we
present abundances of  C, N, O and up to 24 other chemical elements.  Abundances
of carbon were  derived using the ${\rm C}_2$ Swan (0,1) band head at
5635.5~{\AA}. The wavelength interval 7980--8130~{\AA}, with strong
CN features, was analysed in order to determine  nitrogen abundances and
$^{12}{\rm C}/^{13}{\rm C}$  isotope ratios.  The oxygen abundances were
determined from the [O\,{\sc i}] line at 6300~{\AA}.  
Compared with the Sun and
other dwarf stars of the Galactic disk, mean abundances in the investigated
clump stars  suggest that carbon is depleted by about 0.3~dex, nitrogen is
overabundant by about  0.2~dex and oxygen is close to solar. 
This has the effect of lowering the mean C/N ratio to $1.36\pm0.11$.  The mean
$^{12}{\rm C}/^{13}{\rm C}$ ratios are lowered  to  $16\pm2$. Other investigated 
chemical elements have abundance ratios close to the solar ones.
\end{abstract}

\begin{keywords}
stars: abundances -- stars: atmospheres -- stars: horizontal branch -- 
open clusters and associations: individual: IC~4651. 
\end{keywords}

\section{Introduction}

This work is continuing our efforts in studying evolutionary abundance alterations in 
evolved stars of open clusters (Tautvai\v{s}ien\.{e} et al.\ 2000, 2005; Mikolaitis et al.\ 2010).
Our main aim is to determine detailed elemental abundances of carbon,
nitrogen and oxygen, and carbon isotope $^{12}{\rm C}/^{13}{\rm C}$ ratios in 
stars of open clusters in order to better understand reasons of abundance 
alterations caused by stellar evolution. Information on abundances of heavier chemical 
elements will be used for deriving the time evolution of abundances in the Galactic disk 
under the Bologna Open Cluster Chemical Evolution (BOCCE) study 
(Bragaglia \& Tosi 2006, Carretta et al.\ 2007, and references therein).   
In this work, our target of investigations is the open cluster IC~4651. 
    
The open cluster IC\,4651 is an intermediate-age (1.7~Gyr) open cluster 
located 140~pc below the Galactic plane and 7.1~kpc form the Galactic centre (Meibom et al.\ 2002; Pasquini et al.\ 2004).
Meibom et al.\ (2002) provided calculations of the space motion and the Galactic orbit of the cluster. 
The orbital eccentricity is $e = 0.19$ and the mean radius of galactocentric
orbit is 8.6~kpc, its maximum distance from the Galactic plane is 190~pc 
($\alpha_{2000}=17^{h}24.8^{m}, \delta_{2000}=-49^{\circ}56.0^{\prime}; l = 340.088^{\circ}, 
b = -07.907^{\circ}$). 

Results of extensive photometric studies were published first by Eggen (1971) and Lindoff (1972) and later on by 
Anthony-Twarog \& Twarog (1987, 2000), Anthony-Tworag et al.\ (1988). However, the most recent photometric study was performed by
Meibom (2000) and Meibom et al.\ (2002). They combined photometric observations and radial velocity measurements 
for the 44 single member stars down 
to $V = 14.5$~ mag and determined $E(B-V)=0.10$~mag, the distance $d = 1.01 \pm 0.05$~kpc, and the mean radial 
velocity equal to $-30.76 \pm 0.20$~ km\,s$^{-1}$. It was found that 37\% of giant members are spectroscopic
binaries with periods up to 5000~days, and 52\% of the main-sequence and turn-off members are
binaries with periods less than 1000~days. The estimated total mass of IC~4651 is $\approx630 M_{\odot}$ 
(Meibom et al.\ 2002). The turn-off mass of the IC\,4651 stars $M = 1.69M_{\odot}$ was obtained by 
Carretta et al.\ (2004) reading the turn-off values on the Girardi et al.\ (2000) isochrones for solar 
metallicity at the age of the cluster of 1.7~Gyr as determined by Meibom et al.\ (2002). 

There were several photometric studies that evaluated the metallicity of IC\,4651. Based on $uvby-H_{\beta}$ 
photometry, ${\rm [Fe/H]} = 0.23 \pm 0.02$  was found by 
Anthony-Twarog \& Twarog (1987), ${\rm [Fe/H]} = 0.18 \pm 0.05$ by Nissen (1988), and 
${\rm [Fe/H]} = 0.077 \pm 0.012$ by Anthony-Twarog \& Twarog (2000).

High resolution spectroscopic data started to appear in the beginning of 
the millennium. Bragaglia et al.\ (2001) determined the mean metallicity of five evolved stars
 ${\rm [Fe/H]}=0.16 \pm 0.01$. Pasquini et al.\ (2004) provided the cluster metallicity 
${\rm [Fe/H]}= 0.10 \pm 0.03$ from the analysis of 22 faint main sequence stars of the cluster. Abundances of 
the iron peak, $\alpha$-elements and lithium were investigated in their study as well. 
Carretta et al.\ (2004) found the average ${\rm [Fe/H]}=0.11 \pm 0.01$ for {\rm 5 evolved stars of the cluster}, 
which we analyse further in this our work. 
Pace at al.\ (2008) provided the abundance measurements of Fe, Ca, Na, Ni, Ti, Al, Cr, Si for 20 solar-type 
stars belonging to IC\,4651 and found ${\rm [Fe/H]} = 0.12 \pm 0.05$. And finally, Santos et al.\ (2009) derived 
the mean metallicity ${\rm [Fe/H]}=0.15 \pm 0.02$ for IC\,4651 from five dwarf stars. 

In our work for the open cluster IC\,4651, the detailed abundance analysis of almost 30 chemical elements is done. 
Abundances of such key chemical elements as $^{12}{\rm C}$, $^{13}{\rm C}$, N, O as well as representatives 
of s- and r-processes are determined. 

The colour-magnitude diagramme of IC\,4651 with the stars analysed in three most comprehensive chemical abundance 
studies (Pasquini et al.\ 2004; Pace et al.\ 2008 and this work) 
indicated is presented in Fig.~1.

  
\input epsf
\begin{figure}
\epsfxsize=\hsize 
\epsfbox[-5 -10 600 550]{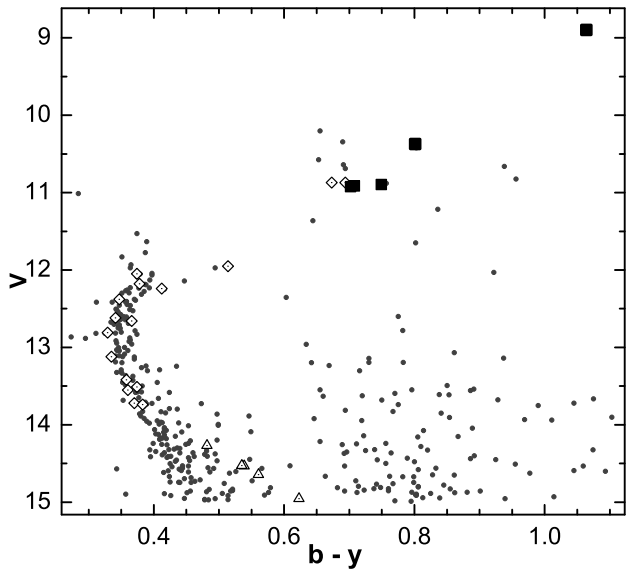} 
\caption{The colour-magnitude diagram of the open cluster IC\,4651. The stars investigated in this work are 
indicated by the filled squares. 
The stars of two other high-resolution spectral abundance studies are shown in this plot as well: the work 
of Pace et al.\ (2008)
is marked by triangles and of Pasquini et al.\ (2004) -- by diamonds.
The diagram is based on Str\"{o}mgren photometry by Anthony-Twarog \& Twarog (2000).} 
\label{fig1}
\end{figure}

\input epsf
\begin{figure}
\epsfxsize=\hsize 
\epsfbox[0 0 500 430]{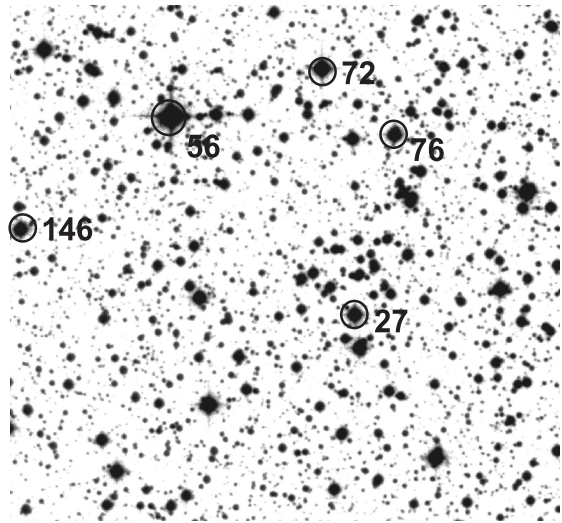} 
\caption{Field of 8 x 8 arcmin$^2$ centered on IC\,4651, with the programme stars indicated 
by their numbers according to Lindoff (1972).}
\label{fig1}
\end{figure}   

\section{Observations and method of analysis}
The spectra of five cluster stars (IC\,4651 27, 56, 72, 76 and 146) were obtained with 
the spectrograph FEROS (Fiber-fed Extended Range Optical 
Spectrograph) mounted at the 1.5~m telescope in La Silla (Chile). The resolving power is 
$R=48\,000$ and the wavelength range is $\lambda\lambda$ 3700--8600\,{\AA}. 
Three stars (27, 76, 146) belong to the red clump of the cluster, the IC 4651 72 star 
is a first-ascent giant, and the star 56 is an RGB-tip giant (see Fig.~1). 
The finding chart of the investigated stars is shown in Fig.~2. The log of observations and 
S/N are presented in the paper by Carretta at al.\ (2004). 

In the same paper by Carretta et al.\ (2004), all the 
main atmospheric parameters for the observed stars were determined. For the convenience 
we present them in this paper as well (Table~1). The effective temperatures were derived 
by minimizing the slope of the abundances from neutral Fe\,{\sc i} lines with respect to the 
excitation potential. Using the line-depth ratios (LDR) technique, Biazzo et al.\ (2007) 
have determined higher effective temperatures for other clump stars in NGC\,4651 and 
have raised doubts that the temperature determinations by Carretta et al. were likely too low. 
We have checked dependences of the other chemical element lines with respect to the excitation 
potential and did not find slopes. Thus, we do not doubt in the correctness of effective temperature 
determinations for IC\,4651 giants by Carretta et al. 
       
The gravities (log~$g$) were derived by Carretta et al. from the iron ionization 
equilibrium. In our study we found a very good agreement between neutral and singly ionised species 
of Cr and Ti, which strongly support the reliability of the atmospheric parameters, in particular the 
gravity values derived from the ionization equilibrium of Fe. The microturbulent velocities were 
determined assuming a relation between 
log~$g$ and $v_t$. The ATLAS models with overshooting were used for the analysis. 
The Fe\,{\sc i}  lines were restricted to the spectral range 5500--7000\,{\AA} in order to minimize 
problems of line crowding and difficulties in the continuum tracing blue-ward. Two examples of 
spectra are presented in Fig.~3. For more details and error estimates, see Carretta et al.

\input epsf
\begin{figure}
\epsfxsize=\hsize 
\epsfbox[20 20 300 220]{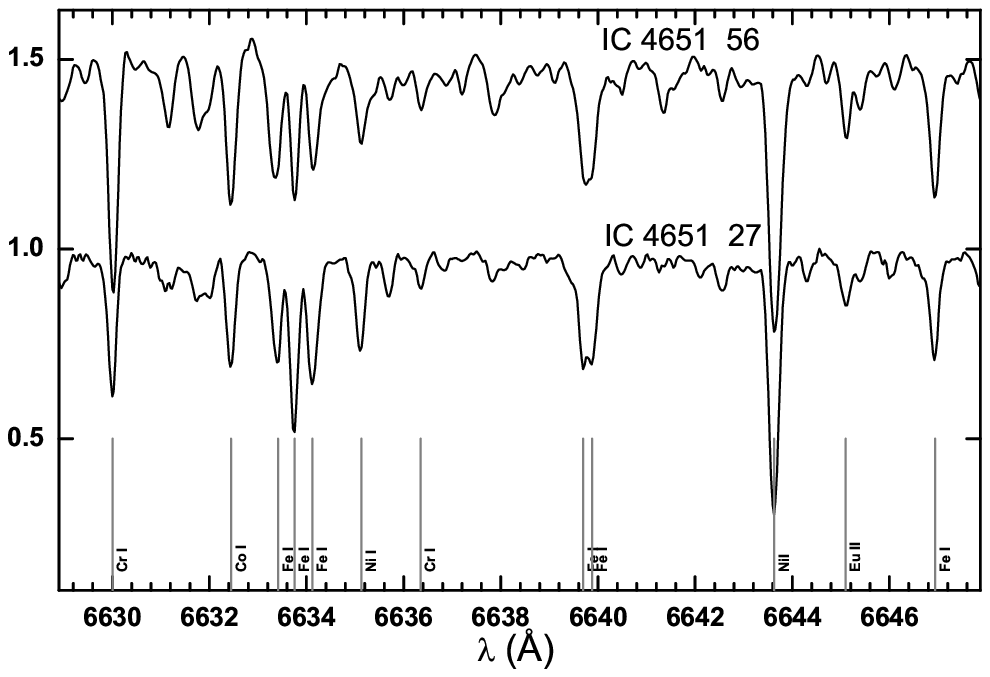} 
    \caption {Small samples of stellar spectra of programme stars in IC\,4651. An offset 
 of 0.5 in relative flux is applied for clarity.}
    \label{CMD}
  \end{figure}

   \begin{table}
\caption[]{Adopted atmospheric parameters for observed stars in IC~4651.
}
\label{Param}
      \[
         \begin{tabular}{rccccrc}
            \hline
            \noalign{\smallskip}
Star$^*$  & $V$ & $B-V$  &  $T_{\rm eff}$ & log~$g$ & [A/H] & $v_t$  \\
         & (mag) & (mag) &     (K)        &       &  & (km s$^{-1}$) \\
\hline
 27 &10.86  & 1.23 & 4610 & 2.52 & 0.10   & 1.17  \\
 56 & 8.95  & 1.68 & 3950 & 0.29 & $-$0.34 & 1.46  \\
 72 &10.41  & 1.33 & 4500 & 2.23 & 0.13   & 1.21  \\
 76 &10.94  & 1.17 & 4620 & 2.26 & 0.11   & 1.21  \\
146 &10.94  & 1.14 & 4730 & 2.14 & 0.10   & 1.21  \\
                \noalign{\smallskip}
            \hline
         \end{tabular}
      \]
$^*$ Star numbers, $V$ and $B-V$  from Lindoff (1972) 
   \end{table}

In this work we used the same method of analysis as in Mikolaitis et al.\ (2010, Paper~I). 
Here we will remind only some details. 

For the ${\rm C}_2$ determination we calculated the 5632 -- 5636~{\AA} interval to compare 
with observations of ${\rm C}_2$ Swan 0 -- 1 band head at 5630.5~{\AA}. 
The interval 7980 -- 8130~{\AA} contains strong $^{12}{\rm C}^{14}{\rm N}$ and $^{13}{\rm C}^{14}{\rm N}$ 
features, so it was used for the nitrogen abundance and $^{12}{\rm C}/^{13}{\rm C}$ ratio analysis. 
We derived the oxygen abundance from synthesis of the forbidden [O\,{\sc i}] line at 6300~{\AA}. 
The $gf$ values for $^{58}{\rm Ni}$ and $^{60}{\rm Ni}$ isotopic line components, which blend the 
oxygen line, were taken from Johansson et al.\ (2003). 
In the spectra of IC\,4651 stars the [O\,{\sc i}] line was not contaminated 
by telluric lines. 

   \begin{table}
      \caption{Effects on derived abundances resulting from model changes 
for the star IC\,4651\,72. The table entries show the effects on the 
logarithmic abundances relative to hydrogen, $\Delta$[A/H]. Note that the 
effects on ``relative" abundances, for example [A/Fe], are often 
considerably smaller than abundances relative to hydrogen, [A/H] } 
        \label{Sens}
      \[
         \begin{tabular}{lrrc}
            \hline
            \noalign{\smallskip}
Species & ${ \Delta T_{\rm eff} }\atop{ +100 {\rm~K} }$ & 
            ${ \Delta \log g }\atop{ +0.3 }$ & 
            ${ \Delta v_{\rm t} }\atop{ +0.3 {\rm km~s}^{-1}}$ \\ 
            \noalign{\smallskip}
            \hline
            \noalign{\smallskip}
C\,(C$_2$)  	&	--0.05	&	0.05	&	0.00	\\
N\,(CN)     	&	0.05	&	0.00	&	0.05	\\
O\,([O\,{\sc i}]) 	&	--0.05	&	--0.05	&	0.00	\\
Na\,{\sc i} 	&	0.09	&	--0.05	&	--0.04	\\
Mg\,{\sc i} 	&	0.03	&	0.00	&	--0.03	\\
Al\,{\sc i} 	&	0.08	&	0.00	&	--0.03	\\
Si\,{\sc i} 	&	--0.06	&	0.07	&	--0.02	\\
Ca\,{\sc i} 	&	0.11	&	--0.04	&	--0.06	\\
Sc\,{\sc ii}	&	--0.02	&	0.13	&	--0.05	\\
Ti\,{\sc i} 	&	0.16	&	--0.02	&	--0.08	\\
Ti\,{\sc ii}	&	--0.03	&	0.13	&	--0.06	\\
V\,{\sc i}  	&	0.16	&	0.01	&	--0.09	\\
Cr\,{\sc i} 	&	0.10	&	0.00	&	--0.06	\\
Cr\,{\sc ii}	&	--0.09	&	0.14	&	--0.03	\\
Mn\,{\sc i} 	&	0.08	&	--0.04	&	--0.06	\\
Co\,{\sc i} 	&	0.02	&	0.05	&	--0.07	\\
Ni\,{\sc i} 	&	0.00	&	0.07	&	--0.05	\\
Cu\,{\sc i} 	&	0.02	&	0.03	&	--0.06	\\
Zn\,{\sc i} 	&	--0.05	&	0.07	&	--0.07	\\
Y\,{\sc ii} 	&	0.16	&	--0.02	&	--0.14	\\
Zr\,{\sc i} 	&	0.00	&	0.12	&	--0.07	\\
Ba\,{\sc ii}	&	--0.03	&	0.14	&	--0.02	\\
La\,{\sc ii}	&	0.03	&	0.08	&	--0.10	\\
Ce\,{\sc ii}	&	0.02	&	0.12	&	--0.03	\\
Nd\,{\sc ii}	&	0.01	&	0.13	&	--0.04	\\
Eu\,{\sc ii}	&	0.03	&	0.13	&	--0.07	\\
$^{12}$C/$^{13}$C&	--2	&	--2	&	1	\\
                 \noalign{\smallskip}
            \hline
         \end{tabular}
      \]
   \end{table}


The abundances of Na and Mg were determined with NLTE taken into 
account as described by Gratton et al.\ (1999). 
Abundances of sodium were determined from equivalent widths of the Na\,{\sc i} lines 
at 5688.22~{\AA}, 6154.23~{\AA} and 6160.75~{\AA}; magnesium from 
the Mg\,{\sc i} lines at 4730.04~{\AA}, 5711.09~{\AA}, 6318.71~{\AA}, 6319.24~{\AA}; 
and aluminum from the Al\,{\sc i} lines at 6696.03~{\AA}, 6698.67~{\AA}, 7835.30~{\AA} and 7836.13~{\AA}.

The determination of zirconium, yttrium, barium, lanthanum,
cerium, neodymium and europium abundances were performed by spectral synthesis method.
The zirconium abundances were derived using the Zr\,{\sc i} lines at 4687~{\AA} and 6127~{\AA}.
We adopted the  barium hyperfine structure and isotopic composition for the 
Ba\,{\sc ii} lines at 5853~{\AA} and 6141~{\AA} from
McWilliam (1998) and for the line at 6496~{\AA} from Mashonkina \& Gehren (2000).
The lanthanum abundances were determined from La\,{\sc ii} lines at 6320~{\AA} and 6390~{\AA}, 
cerium from the Ce\,{\sc ii} lines at 5274~{\AA} and 6043~{\AA}. The neodymium abundance was determined using  
atomic parameters presented by Den Hartog \& Lawler (2003). Due to line crowding in the 
region of neodymium lines, only three Nd\,{\sc ii} lines were chosen: 5092~{\AA}, 5249~{\AA} and 5319~{\AA}. 
The europium abundances were determined using the Eu\,{\sc ii} line at 6645~{\AA}. 
A hyperfine structure for the Eu\,{\sc ii} line was also used for the line synthesis.

\newpage
\begin{landscape}
\begin{table} 
\centering
  \caption{Abundances relative to hydrogen [El/H]. The quoted 
errors, $\sigma$, are the standard deviations in the mean value due to the 
line-to-line scatter within the species. The number of lines used is indicated by $n$. 
The last two columns give the mean [El/Fe] and standard deviations for the cluster stars 
27, 72, 76 and 146.}
\begin{tabular}{lrrrrrrrrrrrrrrrrrrrrrrrrrr}
  \hline
\scriptsize
  & \multicolumn{3}{c}{27} &
  & \multicolumn{3}{c}{56} &
  & \multicolumn{3}{c}{72} &
  & \multicolumn{3}{c}{76} &
  & \multicolumn{3}{c}{146} &
  & \multicolumn{2}{c}{Mean}\\
            \noalign{\smallskip}
\cline{2-4}\cline{6-8}\cline{10-12}\cline{14-16}\cline{18-20}\cline{22-23}
            \noalign{\smallskip}
Species &[El/H] &$\sigma$ &$n$&\ &[El/H] &$\sigma$ &$n$&\ &[El/H] &$\sigma$ &$n$&\ &[El/H] &$\sigma$ &$n$&\ &[El/H] &$\sigma$ &$n$&\ &[El/Fe] & $\sigma$\\ 
            \noalign{\smallskip}
            \hline
            \noalign{\smallskip}
C\,(C$_2$)  	&     --0.15	&	     	&	1	&&	--0.64	&	      	&	1	&&    --0.12	&	      	&	2	&&    --0.17	&	      	&	1	&&     --0.15	&	     	&	1	&&    --0.26	&	0.02	\\
N\,(CN)     	&	0.36	&	0.07	&	24	&&	--0.09	&	0.11	&	18	&&	0.35	&	0.07	&	24	&&	0.30	&	0.06	&	24	&&	0.27	&	0.09	&	23	&&	0.21	&	0.04	\\
O\,([O\,{\sc I}])&	0.05	&	     	&	1	&&	--0.27	&	  	&	1	&&	0.09	&	      	&	1	&&	0.16	&	 	&	1	&&	0.18	&	     	&	1	&&	0.01	&	0.06	\\
Na\,{\sc i} 	&	0.09	&	0.05	&	3	&&	--0.25	&	0.02	&	3	&&	0.10	&	0.04	&	3	&&	0.10	&	0.05	&	3	&&	0.07	&	0.04	&	3	&&	--0.02	&	0.01	\\
Mg\,{\sc i} 	&	0.05	&	0.07	&	4	&&	--0.40	&	0.08	&	4	&&	0.05	&	0.09	&	4	&&	0.09	&	0.05	&	4	&&	0.08	&	0.09	&	4	&&	--0.04	&	0.03	\\
Al\,{\sc i} 	&	0.11	&	0.05	&	4	&&	--0.22	&	0.07	&	4	&&	0.13	&	0.02	&	4	&&	0.17	&	0.04	&	4	&&	0.09	&	0.03	&	4	&&	0.02	&	0.03	\\
Si\,{\sc i} 	&	0.25	&	0.08	&	9	&&	--0.35	&	0.09	&	6	&&	0.25	&	0.10	&	  8	&&	0.20	&	0.07	&	9	&&	0.21	&	0.09	&	9	&&	0.12	&	0.03	\\
Ca\,{\sc i} 	&	0.17	&	0.05	&	9	&&	--0.28	&	0.08	&	6	&&	0.13	&	0.09	&	 10	&&	0.17	&	0.09	&	8	&&	0.17	&	0.10	&	8	&&	0.05	&	0.04	\\
Sc\,{\sc ii}	&	0.17	&	0.02	&	9	&&	--0.50	&	0.05	&	9	&&	0.14	&	0.06	&	  8	&&	0.15	&	0.06	&	9	&&	0.18	&	0.05	&	9	&&	0.05	&	0.03	\\
Ti\,{\sc i} 	&	0.25	&	0.08	&	24	&&	--0.20	&	0.07	&	9	&&	0.23	&	0.09	&	 24	&&	0.18	&	0.08	&	26	&&	0.21	&	0.10	&	28	&&	0.11	&	0.03	\\
Ti\,{\sc ii}	&	0.21	&	0.09	&	6	&&	--0.30	&	0.09	&	8	&&	0.19	&	0.07	&	  7	&&	0.23	&	0.09	&	11	&&	0.14	&	0.09	&	11	&&	0.08	&	0.04	\\
V\,{\sc i}  	&	0.20	&	0.07	&	9	&&	--0.58	&	0.02	&	9	&&	0.17	&	0.09	&	  5	&&	0.25	&	0.05	&	9	&&	0.27	&	0.04	&	9	&&	0.11	&	0.05	\\
Cr\,{\sc i} 	&	0.10	&	0.05	&	19	&&	--0.51	&	0.06	&	11	&&	0.08	&	0.09	&	 25	&&	0.10	&	0.09	&	24	&&	0.10	&	0.09	&	24	&&	--0.01	&	0.02	\\
Cr\,{\sc ii}	&	0.05	&	0.04	&	6	&&	--0.50	&	0.05	&	6	&&	0.00	&	0.09	&	  9	&&	0.08	&	0.09	&	6	&&	0.07	&	0.04	&	6	&&	--0.06	&	0.05	\\
Mn\,{\sc i} 	&	0.20	&	0.04	&	6	&&	--0.47	&	0.01	&	6	&&	0.21	&	0.05	&	  5	&&	0.15	&	0.01	&	6	&&	0.18	&	0.06	&	6	&&	0.07	&	0.03	\\
Co\,{\sc i} 	&	0.25	&	0.07	&	8	&&	--0.18	&	0.09	&	5	&&	0.20	&	0.07	&	  9	&&	0.24	&	0.08	&	8	&&	0.27	&	0.08	&	8	&&	0.13	&	0.04	\\
Ni\,{\sc i} 	&	0.17	&	0.08	&	34	&&	--0.25	&	0.10	&	30	&&	0.20	&	0.09	&	 34	&&	0.17	&	0.07	&	36	&&	0.17	&	0.08	&	36	&&	0.07	&	0.01	\\
Cu\,{\sc i} 	&	0.05	&	0.06	&	3	&&	--0.33	&	0.07	&	3	&&	0.08	&	0.02	&	  3	&&	0.16	&	0.01	&	3	&&	0.10	&	0.02	&	3	&&	--0.01	&	0.05	\\
Zn\,{\sc i} 	&	0.00	&	0.06	&	2	&&	--0.28	&	0.02	&	2	&&	0.10	&	0.09	&	  2	&&	0.07	&	0.07	&	2	&&	0.07	&	0.08	&	2	&&	--0.05	&	0.03	\\
Y\,{\sc ii} 	&	0.07	&	0.05	&	6	&&	--0.25	&	0.03	&	6	&&	0.10	&	0.04	&	6	&&	0.04	&	0.01	&	6	&&	0.14	&	0.04	&	6	&&	--0.02	&	0.05	\\
Zr\,{\sc i} 	&	0.03	&	0.08	&	2	&&	--0.34	&	0.01	&	2	&&	0.00	&	0.05	&	2	&&	0.04	&	0.02	&	2	&&	0.00	&	0.05	&	2	&&	--0.09	&	0.03	\\
Ba\,{\sc ii}	&	0.03	&	0.03	&	3	&&	--0.29	&	0.04	&	2	&&	0.08	&	0.07	&	3	&&	0.05	&	0.05	&	3	&&	0.05	&	0.05	&	3	&&	--0.06	&	0.01	\\
La\,{\sc ii}	&	0.18	&	0.02	&	2	&&	--0.33	&	0.02	&	2	&&	0.10	&	0.05	&	2	&&	0.15	&	0.03	&	2	&&	0.15	&	0.05	&	2	&&	0.03	&	0.05	\\
Ce\,{\sc ii}	&	0.22	&	0.08	&	2	&&	--0.30	&	0.02	&	2	&&	0.20	&	0.07	&	2	&&	0.24	&	0.01	&	2	&&	0.20	&	0.07	&	2	&&	0.11	&	0.03	\\
Nd\,{\sc ii}	&	0.27	&	0.08	&	3	&&	--0.18	&	0.04	&	2	&&	0.28	&	0.06	&	3	&&	0.15	&	0.05	&	3	&&	0.18	&	0.03	&	3	&&	0.11	&	0.06	\\
Eu\,{\sc ii}	&	0.14	&	     	&	1	&&	--0.25	&	      	&	1	&&	0.10	&	      	&	1	&&	0.13	&	      	&	1	&&	0.10	&	     	&	1	&&	0.01	&	0.03	\\
\\              		      		     		   		       		      		    		      		      		  		      		      		   		       		     		   		      		    	
C/N             &	1.23	&	     	&	   	&&	1.12	&	      	&	    	&&	1.34	&	      	&	  	&&	1.35	&	      	&	   	&&	1.5	&	     	&	   	&&	1.36	&	0.11	\\
$^{12}$C/$^{13}$C&	17	&	     	&	   	&&	14	&	      	&	    	&&	15	&	      	&	  	&&	14	&	      	&	   	&&	18	&	     	&	   	&&	16	&	2	\\

             \noalign{\smallskip}   
\hline
         \end{tabular} 
\end{table}
\end{landscape}      

\subsection{Estimation of uncertainties}

The sources of uncertainty were described in detail in our Paper~I. 

The sensitivity of the abundance 
estimates to changes in the atmospheric parameters by the assumed errors 
($\pm~100$~K for $T_{\rm eff}$, $\pm 0.3$~dex for log~$g$ and 
$\pm 0.3~{\rm km~s}^{-1}$ for $v_{\rm t}$) is 
illustrated  for the star IC\,4651 72 (Table~2). It is seen that possible 
parameter errors do not affect the abundances seriously; the element-to-iron 
ratios, which we use in our discussion, are even less sensitive. 
The sensitivity of iron abundances to stellar atmospheric parameters were described 
in Carretta et al.\ (2004). 

The scatter of the deduced line abundances $\sigma$, presented in Table~3, 
gives an estimate of the uncertainty due to the random errors, e.g. in 
continuum placement and the line parameters (the mean value of  $\sigma$ 
is $0.06$). Thus the uncertainties in the derived abundances that are the 
result of random errors amount to approximately this value.

Since abundances of C, N and O are bound together by the molecular equilibrium 
in the stellar atmosphere, we have also investigated how an error in one of 
them typically affects the abundance determination of another. 
$\Delta{\rm [O/H]}=0.10$ causes 
$\Delta{\rm [C/H]}=0.05$ and $\Delta{\rm [N/H]}=-0.10$,   
$\Delta{\rm [C/H]}=0.10$ causes $\Delta{\rm [N/H]}=-0.15$ and 
$\Delta{\rm [O/H]}=0.02$, 
$\Delta {\rm [N/H]}=0.10$ has no effect on either the carbon or the oxygen abundances.

\section{Results and discussion}

The abundances relative to hydrogen
[El/H]\footnote{In this paper we use the customary spectroscopic notation
[X/Y]$\equiv \log_{10}(N_{\rm X}/N_{\rm Y})_{\rm star} -
\log_{10}(N_{\rm X}/N_{\rm Y})_\odot$} and $\sigma$ (the line-to-line 
scatter) derived for up to 27 neutral and ionized 
species (including $^{13}$C) for the 
programme stars are listed in Table~3.
The average cluster abundances [El/Fe] and dispersions about the mean values for IC\,4651 are 
presented in Table~3 as well. They are calculated from the results determined for the stars 27, 72, 76 and 146. 
Due to the different [Fe/H], which is by 0.4~dex lower than of other cluster 
stars, the star 56 was not used in the average calculations even though its values 
do not change the average abundances much. For the majority of the chemical elements the changes are just 
$\pm 0.01-0.02$~dex, only for Si\,{\sc i} and Sc\,{\sc ii} the difference is 0.04~dex, and for V\,{\sc i} -- 0.07~dex. 
From its element to iron ratios, C/N and  $^{12}{\rm C}/^{13}{\rm C}$ ratios, this star within errors of uncertainties 
is indistinguishable from other evolved stars of this cluster. However, we think that with more adequate model 
atmospheres this star can be investigated more accurately.
In Fig.~4 we show an example of spectrum syntheses for the [O\,{\sc i}] line in IC\,4651\,56. 
In Fig.~3 a sample of its spectrum is shown as well.

In IC\,4651, the majority of investigated 
chemical elements have abundance ratios close to the solar ones. 
The mean cluster [$\alpha/{\rm Fe}] \equiv {1\over 4}
([{\rm Mg}/{\rm Fe}]+[{\rm Si}/{\rm Fe}]+[{\rm Ca}/{\rm Fe}]
+[{\rm Ti}/{\rm Fe}]) = 0.06\pm0.07$~(s.d.), which is close to the solar value. 

Abundances of Na\,{\sc i}, Al\,{\sc i}, Si\,{\sc i}, Ca\,{\sc i}, Ti\,{\sc i} and Ni\,{\sc i} were 
determined for the main-sequence stars of IC\,4651 by Pace et al.\ (2008). The mean [El/Fe] ratios 
in these stars are very close to solar as well.  

Pasquini et al.\ (2004) investigated spectra of both giants and main-sequence stars in IC\,4651. 
The authors expressed their strong believe that [Na/Fe] ratio is comprehensively 
higher in the giants in comparison to the main-sequence stars and that this is due to internal nucleosynthesis and 
mixing. However, neither Pace et al.\ (2008), neither our study may confirm this statement. In Table~3, 
we present the mean [El/Fe] for giant and main-sequence stars of IC\,4651 investigated in our work, 
Pasquini et al.\ (2004) and Pace et al.\ (2008). In our work, the abundances of Na and Mg were determined with NLTE taken 
into account, and we do not find an overabundance of these chemical elements. 
         
\input epsf
\begin{figure}
\epsfxsize=\hsize 
\epsfbox[-20 -20 620 470]{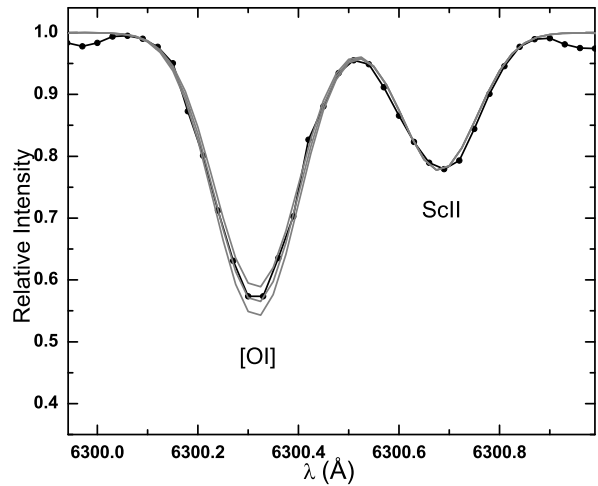} 
    \caption {Fit to the forbidden [O\,{\sc i}] line at 6300~{\AA} in 
IC\,4651\ 56. The observed spectrum is shown as a solid line with black dots. Synthetic
spectra with [O/Fe]$=0.3$, $0.08$, and $0.13$ are shown as solid gray lines.}
    \label{Oxygen}
  \end{figure}

In Fig.~4 of Paper~I, we presented the radial distribution of some elemental abundances for 
BOCCE clusters analysed so far, and for others in recent studies. 
IC\,4651 agrees well with results 
of other open clusters at the same $R_{\rm gc}$ of 7.1~kpc.   

\begin{table} 
\centering
  \caption{The mean [El/Fe] for giant (G) and main-sequence (MS) stars of IC\,4651 investigated in this work, 
Pasquini et al.\ (2004) and Pace et al.\ (2008).}
\begin{tabular}{lccccccc}
  \hline
\scriptsize
  & \multicolumn{1}{c}{This work }&
  & \multicolumn{2}{c}{Pasquini } &
  & \multicolumn{1}{c}{Pace } \\
Species &[El/Fe]&\ &[El/Fe]&[El/Fe]&\ &[El/Fe] \\ 
        &   G   &\  &   G   &   MS  &\  &MS    \\ 
            \hline
            \noalign{\smallskip}
Na\,{\sc i} 	&	0.00	& &0.19 &--0.09 & &--0.03\\
Mg\,{\sc i} 	&	--0.05	& &0.09 &  0.13&  &\\
Al\,{\sc i} 	&	0.03	& &0.07 &--0.07 & &--0.10\\
Si\,{\sc i} 	&	0.09    & &0.08 & 0.07 & &--0.02\\
Ca\,{\sc i} 	&	0.05	& &0.00 & 0.04& &0.04\\
Sc\,{\sc ii}	&	0.01	& &0.11 &--0.11& & \\
Ti\,{\sc i} 	&	0.11	& &0.12 &  0.08 & &--0.02\\
Ti\,{\sc ii}	&	0.07	& &0.18 & 0.00&  & \\
Cr\,{\sc i} 	&	--0.05	& &--0.02 &0.11 & & \\
Ni\,{\sc i} 	&	0.07	& & 0.10 & 0.01 & &--0.02\\
             \noalign{\smallskip}   
\hline
         \end{tabular}
\end{table}

In the following sections 
we will discuss in more detail the results of carbon and nitrogen abundance 
determinations.  

\input epsf
\begin{figure}
\epsfxsize=\hsize 
\epsfbox[-20 -20 620 470]{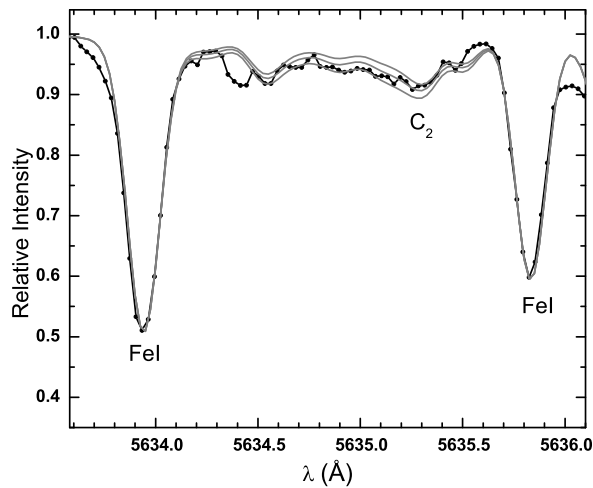} 
    \caption {Small region of IC\,4651 27 spectrum (solid black line with black dots) at 
${\rm C}_2$ Swan (0,1) band head 5635.5~{\AA}, plotted together with 
synthetic spectra with [C/Fe] values lowered by $-0.2$~dex (lower gray line), $-0.25$~dex (middle gray line) 
and $-0.3$ (upper gray line). 
}
    \label{}
  \end{figure}

\input epsf
\begin{figure}
\epsfxsize=\hsize 
\epsfbox[0 0 620 470]{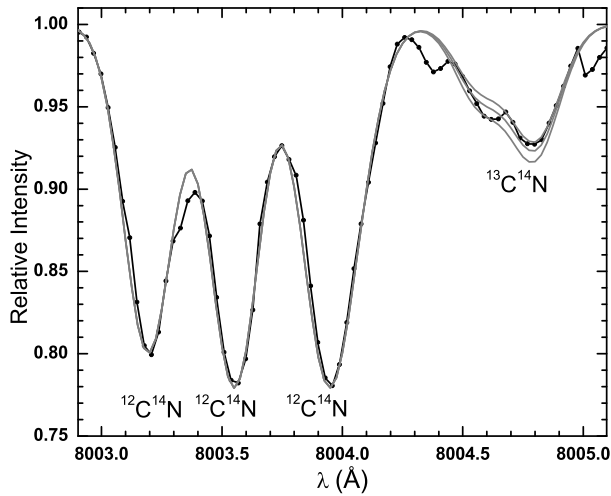} 
    \caption {Small region of IC\,4651\ 72 spectrum (solid black line with black dots) 
    with $^{13}{\rm C}^{14}{\rm N}$ feature.
Grey lines show synthetic spectra with $^{12}{\rm C}$/$^{13}{\rm C}$ ratios equal to 13 (lower line), 
15 (middle line) and 17(upper line). 
}
    \label{Oxygen}
  \end{figure}
\subsection{Carbon and nitrogen abundances}

The average value of carbon to iron ratio in IC\,4651 is ${\rm [C/Fe]}=-0.27\pm0.02$. 
In Fig.~5, a fit to the IC\,4651\ 27 spectrum at ${\rm C}_2$ 5635.5~{\AA} is
shown.  

We compared the carbon abundance in IC\, 4651 with carbon abundances determined for dwarf stars 
in the Galactic disk. Shi et al.\ (2002) performed an abundance 
analysis of carbon for a sample of 90 F and G type main-sequence disk stars  
using C\,{\sc i} and [C\,{\sc i}] lines and found [C/Fe] to be about solar 
at the solar metallicity.   
Roughly solar carbon abundances were found by Gustafsson et al.\ (1999) 
who analysed a sample of 80 late F and early G type dwarfs using the forbidden 
[C\,{\sc i}] line. 
The ratios of [C/Fe] in our stars lie about 0.3~dex below 
the values obtained for dwarf stars of the Galactic disk.      

The mean nitrogen to iron abundance ratio 
in IC\,4651 is ${\rm [N/Fe]}=0.21\pm0.04$. 
This shows that nitrogen is overabundant in these evolved stars of IC\,4651,
since [N/Fe] values in the Galactic main-sequence stars are about solar at the solar
metallicity (c.f. Shi et al.\ 2002). Unfortunately, neither carbon, neither nitrogen 
abundances were investigated in the main sequence stars of IC\,4651
by Pace et al.\ (2008) and Pasquini et al.\ (2004).   

The mean C/N ratios in IC\,4651 is equal to $1.36\pm 0.11$. 
The smallest value of ${\rm C/N}=1.12$ was obtained for the star IC\,4651\,56. 

The $^{12}{\rm C}/^{13}{\rm C}$ ratios were determined for all programme stars from the (2,0)
$^{13}{\rm C}^{14}{\rm N}$ feature at 8004.728~{\AA}. In Fig.~6 we show a small region of 
IC\,4651\,72 spectrum together with spectral syntheses obtained with three different values of the carbon isotopic ratio.
We find that the mean $^{12}{\rm C}/^{13}{\rm C}$ ratios are about  $16\pm2$ in the evolved stars investigated. 
   
The solar carbon and nitrogen abundances used in our work are log$A_{\rm C} = 8.52$ and 
log$A_{\rm N} = 7.92$ (Grevesse \& Sauval 2000), so the solar ${\rm C/N} = 3.98$.
The $^{12}{\rm C}/^{13}{\rm C}$ ratio in the solar photosphere is equal to 89 
(Coplen et al.\ 2002).  
  
\subsection{Comparison of $^{12}{\rm C}/^{13}{\rm C}$ and C/N ratios with theoretical models}

\begin{table*}
\begin{center}
\caption{$^{12}{\rm C}/^{13}{\rm C}$ and C/N ratios along with turn-off mass, age, galactocentric distance and 
atmospheric parameters for clump stars}
\begin{tabular}{lrcccccccccc}
\hline
Cluster & Star  & $M_{\rm TO}$($M_{\odot}$) & Age(Gyr) & $R_{\rm gc}$(kpc)  &  T$_{\rm eff}$(K) & $\log$ $g$ & [A/H] & $^{12}{\rm C}/^{13}{\rm C}$ & C/N & Ref.$^*$ \\
\hline
 NGC\,752&   1  &1.60&2.0&8.75&5000 & 2.85 &  0.1   & 16 & -- & 1\\
        &   75  &   &  & &4900 & 2.85 &  0.1   & 13 & -- & 1\\
        &   77  &   &  & &4900 & 2.85 &  0.2   & 16 & -- & 1\\
        &  213  &   &  & &5000 & 2.90 &  0.1   & 14 & -- & 1\\
        &  295  &   &  & &5000 & 2.90 &  0.2   & 15 & -- & 1\\
 \noalign{\smallskip}
NGC\,2360 & 50 & 2.02  & 1.15  & 6.32&5015    &2.90   &--0.03   &--     &1.04 & 5\\
	  & 62 &       &       &      &5105    &3.15   & 0.12    &--     &1.38 & 5\\
	  & 86 &       &       &      &4960    &2.65   &--0.06   &--     &0.93 & 5\\
 \noalign{\smallskip}
          & 12    & & &              &4800 & 2.70 &  0.2   & 14.5 & -- & 1\\
 \noalign{\smallskip}
NGC\,2447 & 28  & 1.90  & 0.45   &6.51 &5060    &2.70   &--0.01   &--   &0.69 & 5\\
	  & 34  &       &       &      &5120    &2.90   &--0.01   &--   &0.87 & 5\\
 \noalign{\smallskip}
NGC\,2682 & F84   & 1.20  & 5.0 &9.05 &4750 & 2.4 & --0.02   & 20 &  1.15 & 3\\
          & F141   &   &  & &4730 & 2.4 & --0.01   & 16 &  1.32 & 3\\
          & F151   &   &  & &4760 & 2.4 & --0.03   & 17 &  1.32 & 3\\
          & F164   &   &  & &4700 & 2.5 &   0.00   & 18 &  1.62 & 3\\
          & F224   &   &  & &4710 & 2.4 & --0.11   &  8 &  1.58 & 3\\
          & F226   &   &  & &4730 & 2.4 & --0.02   & 15 &  1.62 & 3\\
            \noalign{\smallskip}
          & F84  & &  & &4800 & 2.70 &  0.0   & 11.5 & -- & 1\\
          & F141 & & & &4800 & 2.70 &  0.0   & 10.5 & -- & 1\\
          & F164 & &  & &4800 & 2.70 &  0.0   & 10.5 & -- & 1\\
 \noalign{\smallskip}
NGC\,2714 & 5 	&2.91	&0.40	& 8.34 &5070 	&2.70 	&0.12	&-- 	&0.83 & 5\\
 \noalign{\smallskip}
NGC\,3532 & 19 	&3.03	&0.35	& 7.87 &4995 	&2.65 	&0.11	&12 	&1.02 & 5\\
	  & 122 &       &       &      &5045    &2.60   &--0.02   &--     &0.93 & 5\\
	  & 596 &       &       &      &5020    &2.50   &0.04   &--     &0.95 & 5\\
           \noalign{\smallskip}
	  & HD95879  & &  & &5000 & 2.25 &  0.08   & 10 & 1.10 & 2\\
	  & HD96174  & &  & &5000 & 2.17 &  0.00   & 15 & 0.59 & 2\\
	  & HD96175  & &  & &5100 & 2.25 &  0.12   & 15 & 0.44 & 2\\
	  & HD96445  & &  & &5000 & 2.36 &  0.13   & 10 & 1.51 & 2\\
 \noalign{\smallskip}
NGC\,5822 & 201 	&2.19	&0.9	& 8.10 &5035 	&2.85 	&0.05	&13 	&0.87 & 5\\
          & 316 	&  	& 	&      &5110 	&3.05 	&0.16	&-- 	&1.00 & 5\\
 \noalign{\smallskip}
NGC\, 6134 & 39 & 2.34 & 0.7 & 7.6 & 4980 & 2.52 & 0.24 &  9 & 1.48 & 6\\   
           & 69 &  & & & 4950 & 2.83 & 0.11 & 12 & 1.38 & 6\\
           & 75 &  & & & 5000 & 3.10 & 0.22 &  7 & 1.41 & 6\\
           & 114 &  & & & 4940 & 2.74 & 0.11 &  6 & 1.05 & 6\\    
           & 129 &  & & & 5000 & 2.98 & 0.05 &  8 & 0.98 & 6\\
           & 157 & & & & 5050 & 2.92 & 0.16 & 12 & 1.10 & 6\\
            \noalign{\smallskip}
           & 30 & 	&   &    &4980 	&2.95 	&0.21	&12 	&0.93 & 5\\
 \noalign{\smallskip}
NGC\,6281 & 3 	& 3.18	&0.3	& 8.47 &4915 	&2.30 	&0.01	&12 	&0.64 & 5\\
          & 4 	& 	&	&      &5015 	&2.50 	&0.09	&12 	&0.95 & 5\\
 \noalign{\smallskip}
NGC\,6633 & 100 & 2.79	& 0.45	& 8.42 &5015 	&2.85 	&0.11	&21 	&0.91 & 5\\
 \noalign{\smallskip}
NGC\,7789 & K605   & 1.60 & 1.4 & 9.43 & 4860 & 2.4 & --0.02   & 10 &  1.05 & 4\\
          & K665   &      &     &      & 4970 & 2.4 &  0.00   &  9 &  1.45 & 4\\
          & K732   &      &     &      & 4900 & 2.3 &  0.02   &  7 &  1.51 & 4\\
 \noalign{\smallskip}
IC\,2714 & 5 	&2.85	&0.40	& 8.34 &5070 	&2.70 	&0.12	&-- 	&0.83 & 5\\
 \noalign{\smallskip}
IC\,4651 & 27 &1.69 & 1.7 &7.1 &4610 & 2.52 & 0.10   & 17 & 1.23 & 7\\
         & 76 &     &     &     &4620 & 2.26 & 0.11   & 14 & 1.35 & 7\\
         & 146&     &     &     &4730 & 2.14 & 0.10   & 18 & 1.50 & 7\\
 \noalign{\smallskip}
IC\,4756 & 12 	& 2.37	& 0.7	& 7.23 &5030 	&2.75 	&--0.01	&11 	&0.91 & 5\\
         & 14 	&       &	&      & 4720 	&2.47 	&0.03	&17 	&1.02 & 5\\
         & 38 	& 	&	&      & 5075 	&3.00 	&0.05	&10 	&1.20 & 5\\
         & 69 	& 	&	&      & 5130 	&3.00 	&0.08	&5 	&1.15 & 5\\
            \noalign{\smallskip}
        & 144   &   &  & &5200 & 3.20 &  0.0   & 18 & -- & 1\\
        & 176   &   &  & &5200 & 3.00 &  0.0   & 12 & -- & 1\\
        & 228   &   &  & &5000 & 2.90 &  0.0   & 21 & -- & 1\\
        & 296   &   &  & &5000 & 2.90 &  0.0   & 18 & -- & 1\\
\hline
\end{tabular}
\label{CCdata}
\end{center}
$^*$ 1 -- Gilroy (1989); 2 -- Luck(1994); 3 -- Tautvai\v{s}ien\.{e} et al.\ (2000); 4 -- Tautvai\v{s}ien\.{e} et al.\ (2005); 
5 -- Smiljanic et al.\ (2009);\\
 6 -- Mikolaitis et al.\ (2010);  7 -- This work.
\end{table*}

The carbon and nitrogen abundances, C/N and especially the carbon isotope 
ratios $^{12}{\rm C}/^{13}{\rm C}$ are key tools for stellar evolution studies.  
Investigations of abundances of these chemical elements in atmospheres of clump stars 
of open clusters may provide a comprehensive information on 
chemical composition changes. 
The clump stars have accumulated all chemical composition changes that have happened during 
their evolution along the giant branch and during the helium flash.  
 
In Fig.~7 and 8, we compare the mean carbon isotope and C/N ratios of clump stars in different open clusters as 
a function of turn-off mass with the theoretical models of the $1^{st}$ dredge-up, thermohaline mixing (TH), 
thermohaline mixing together with rotation-induced mixing for stars at the zero age main sequence (ZAMS) having 
rotational velocities of 110\,km\,s$^{-1}$, 250\,km\,s$^{-1}$ and 300\,km\,s$^{-1}$ computed by 
Charbonnel \& Lagarde (2010); and Cool Bottom Processing model (CBP) by Boothroyd \& Sackman (1999).

The most resent modelling of extra-mixing processes was done by Charbonnel \& Lagarde (2010). 
They are based on ideas of Eggleton et al.\ (2006) and Charbonnel \& Zahn (2007). Eggleton et al.\ (2006) found 
a mean molecular weight ($\mu$) inversion in their $1 M_{\odot}$ stellar evolution model, occurring after 
the so-called luminosity bump on the red giant branch, when the H-burning shell source enters the chemically 
homogeneous part of the envelope. The $\mu$-inversion is produced by the reaction 
$^3{\rm He(}^3{\rm He,2p)}^4{\rm He}$, as predicted by Ulrich (1972). It does not occur earlier, because the 
magnitude of the $\mu$-inversion is small and negligible compared to a stabilizing 
$\mu$-stratification. Following Eggleton et al.\ (2006), Charbonnel \& Zahn (2007) have computed stellar models 
including the prescription by Ulrich (1972) and extended them to the case of a non-perfect gas for the turbulent 
diffusivity produced by that instability in a stellar radiative zone. They found that a double diffusive instability 
referred to as thermohaline convection, which has been discussed long ago in the literature (Stern 1960), 
is important in evolution of red giants. This mixing connects the convective envelope with the external wing of 
hydrogen burning shell and induces surface abundance modifications in red giant stars. 

Charbonnel \& Lagarde (2010)
also computed the models of rotation-induced mixing for stars at the zero age main sequence having 
rotational velocities of 110\,km\,s$^{-1}$, 250\,km\,s$^{-1}$ and 300\,km\,s$^{-1}$. Typical initial ZAMS rotation 
velocities were chosen depending on the stellar mass based on observed rotation distributions in young open clusters 
(Gaig\'{e} 1993). The convective envelpoe was supposed to rotate as a solid body through the evolution. The transport 
coefficients for chemicals associated to thermohaline and rotation-induced mixings were simply added in the diffusion 
equation and the possible interactions between the two mechanisms were not considered. The rotation-induced mixing 
modifies the internal chemical structure of main sequence stars, although its signatures are revealed only later in 
the stellar evolution.   

The models by Boothroyd \& Sackmann (1999) include the deep circulation mixing below the base of 
the standard convective envelope, and the consequent "cool bottom processing" (CBP) of CNO isotopes.      

The theoretical models were compared to the observational data of $^{12}{\rm C}/^{13}{\rm C}$ and C/N listed 
in Table~5, which we collected for clump stars of open clusters 
investigated by Mikolaitis et al.\ (2010),  Smiljanic et al.\ (2009), Tautvai\v{s}ien\.{e} 
et al.\ (2000, 2005), Luck (1994) and Gilroy (1989). From Gilroy (1989) we selected 4 clusters with well defined 
red clump stars. Luck (1994) derived carbon isotope ratios for 8 open 
clusters, however only one cluster was included to our comparison since for other clusters it was very difficult 
to identify stars of red clump. 
The turn-off masses, ages and galactocentric distances were choosen from the most recent studies and used for displaying of 
other $^{12}{\rm C}/^{13}{\rm C}$ and C/N  investigations for the same cluster, if available. 
 
In Fig.~7 and 8 we can see that for clusters with stars of smaller turn-off masses the $^{12}{\rm C}/^{13}{\rm C}$ and C/N values are 
in a good agreement with both models of extra-mixing used for the comparison. However, $^{12}{\rm C}/^{13}{\rm C}$ values in the clump stars 
with turn-off masses of 2--3~$M_{\odot}$ in most of the investigated clusters are lower than predicted by 
the available models and need modelling of larger extra-mixing.

\input epsf
\begin{figure}
\epsfxsize=\hsize 
\epsfbox[-20 -20 620 500] {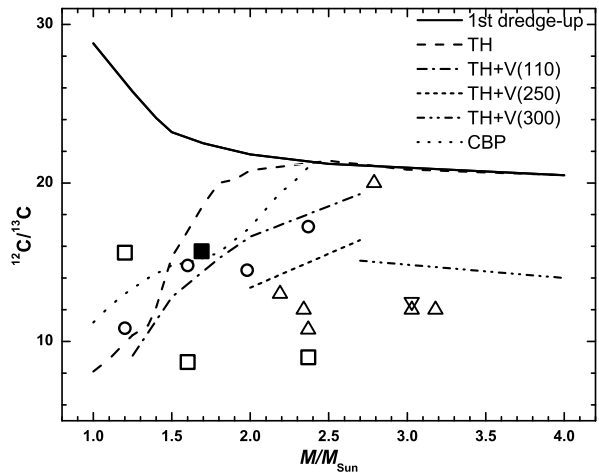} 
    \caption {The average carbon isotope ratios in clump stars of open clusters
as a function of stellar  turn-off mass. 
The result of this work is marked by the filled square; 
from Mikolaitis et al.\ (2010) and Tautvai\v{s}ien\.{e} et al.\ (2000, 2005) -- open squares; 
from Smiljanic et al.\ (2009) -- open triangles; from Luck (1994) -- reversed open triangle;
from Gilroy (1989) -- open circles. 
The models of the $1^{st}$ dredge-up, thermohaline mixing (TH) and rotation-induced 
mixing (V) are taken from Charbonnel \& Lagarde (2010). The
CBP model of extra-mixing is taken from Boothroyd \& Sackmann (1999). 
}
    \label{C/C}
  \end{figure}

\input epsf
\begin{figure}
\epsfxsize=\hsize 
\epsfbox[-20 -20 620 500] {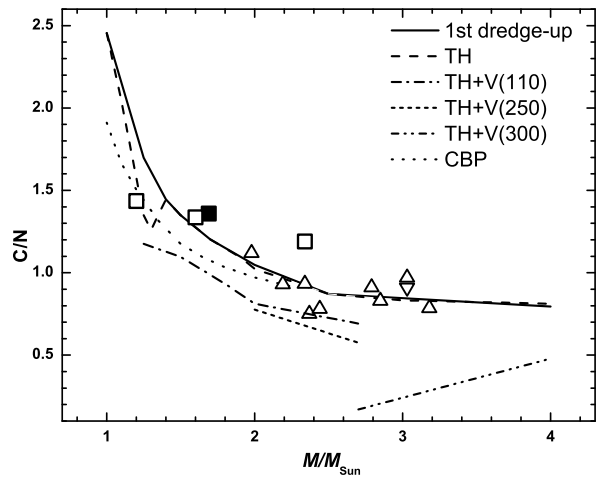} 
    \caption {The average carbon to nitrogen ratios in clump stars of open clusters
as a function of stellar  turn-off mass. The meaning of symbols are as in Fig.~7.
}
    \label{C/N}
  \end{figure}

\subsection{Final remarks}

Carbon and nitrogen are 
important products of nucleosynthesis processes in stellar interiors, and the evidence of 
their abundance variation during stellar evolution is a signature of physical mixing processes 
between the atmosphere and deeper layers of a star.
Such abundance alterations may be well traced in open clusters. They provide a unique possibility for 
investigation of a number of stars of nearly the same age, 
distance and origin, as open cluster stars are claimed to be formed in the same protocloud of gas and dust.
Open clusters have a high reliability of mass, distance, evolutionary phase and 
abundance determinations. 

Extra-mixing processes may become efficient on the red giant branch when 
stars reach the so-called RGB  bump, and may modify the surface 
abundances. It is known that alterations of $^{12}{\rm C}/^{13}{\rm C}$ and 
$^{12}{\rm C}/^{14}{\rm N}$ ratios depend on stellar evolutionary stage, mass and 
metallicity (see Charbonnel et al.\ 1998, Gratton et al.\ 2000, Chanam\'{e} et al.\ 2005, 
Cantiello \& Langer 2010; Charbonnel \& Lagarde 2010 for more discussion). 
 
The comparison of the observational data
with theoretical models of stellar evolution shows that processes of extra-mixing in stars of 
open clusters with turn-off masses of 2--3~$M_{\odot}$ are larger than predicted.

\section*{Acknowledgments}

This research has made use of Simbad, VALD and NASA ADS databases.
Bertrand Plez (University of Montpellier II) and Guillermo Gonzalez  
(Washington State University) were particularly generous in providing us with 
atomic data for CN and C$_2$ molecules, respectively.
\v{S}.\,M. and G.\,T. were supported by the Ministry of Education and Science of Lithuania 
via LitGrid programme and by the European Commission via FP7 Baltic Grid II project. 

    

\begin{thebibliography}{}
 \bibitem [\protect\citeauthoryear{Anthony-Twarog \& Twarog}{1987}]{Anthony-Twarog1987} Anthony-Twarog B. J. \& Twarog B. A., 1987, AJ, 94, 1222
\bibitem [\protect\citeauthoryear{Anthony-Twarog \& Twarog}{2000}]{Anthony-Twarog2000} Anthony-Twarog B. J. \& Twarog B. A., 2000, AJ, 119, 2282 
\bibitem [\protect\citeauthoryear{Anthony-Twarog et al.}{1988}]{Anthony-Twarog1988} Anthony-Twarog B. J., Mukherjee K., Twarog B. A., Caldwell N., 1988, AJ, 95, 1453
 \bibitem [\protect\citeauthoryear{Anthony-Twarog \& Twarog}{2000}]{Anthony-Twarog2000} Anthony-Twarog B. J., Twarog B. A., 2000, AJ, 119, 2282
 \bibitem [\protect\citeauthoryear{Boothroyd et al.}{1999}]{Boothroyd1999} Boothroyd A. I., Sackmann I. J., 1999, ApJ, 510, 232
 \bibitem [\protect\citeauthoryear{Biazzo et al.}{2007}]{Biazzo2007} Biazzo K., Pasquini L., Girardi L., Frasca A., da Silva L., Setiawan, J., Marilli, E., Hatzes, A. P., Catalano, S.
 \bibitem [\protect\citeauthoryear{Bragaglia et al.}{2006}]{Bragaglia2006} Bragaglia A., Tosi M. 2006, AJ, 131, 1544
 \bibitem [\protect\citeauthoryear{Bragaglia et al.}{2001}]{Bragaglia2001} Bragaglia A., et al.  2001, AJ, 121, 327
 \bibitem [\protect\citeauthoryear{Cantiello \& Langer}{2010}]{Cantiello2010} Cantiello M. \& Langer N., 2010, A\&A, 521, 9
 \bibitem [\protect\citeauthoryear{Carretta et al.}{Carretta2007}]{Carretta2007} Carretta  E., Bragaglia A., Gratton R. 2007, A\&A, 473, 129
 \bibitem [\protect\citeauthoryear{Carretta et al.}{2004}]{Carretta2004} Carretta  E., Bragaglia A., Gratton R., Tosi M., 2004, A\&A, 422, 951
 \bibitem [\protect\citeauthoryear{Chanam\'{e} et al.}{2005}]{Chaname2005} Chanam\'{e} J., Pinsonneault M., Terndrup D. M., 2005, ApJ, 631, 540
 \bibitem [\protect\citeauthoryear{Charbonnel et al.}{1998}]{Charbonnel1998} Charbonnel C., Brown J. A., Wallerstein G., 1998, A\&A 332, 204
 \bibitem [\protect\citeauthoryear{Charbonnel \& Zahn}{2007}]{Charbonnel2007} Charbonnel, C. \& Zahn, J.-P. 2007, A\&A, 467, 15
 \bibitem [\protect\citeauthoryear{Charbonnel \& Lagarde}{2010}]{Charbonnel2010} Charbonnel C. \& Lagarde N., 2010, A\&A, 522, 10
 \bibitem [\protect\citeauthoryear{Coplen et al.}{2002}]{Coplen2002} Coplen T.\ et al.\ 2002, Pure \& Appl.\ Chem.\ 74:1987-2017 
 \bibitem [\protect\citeauthoryear{Den Hartog et al.}{2003}]{DenHartog2003} Den Hartog E. A., Lawler J. E., Sneden C., Cowan J. J., 2003, ApJS, 148, 543
 \bibitem [\protect\citeauthoryear{Eggen}{1971}]{Eggen1971} Eggen, O. J., 1971, ApJ, 166, 87
 \bibitem [\protect\citeauthoryear{Eggleton}{2006}]{Eggleton2006} Eggleton P. P., Dearborn D. S. P., Lattanzio J. C., 2006, Sci, 314, 1580
 \bibitem [\protect\citeauthoryear{Gaig\'{e}}{1993}]{Gaige1993}Gaig\'{e} Y., 1993, A\&A, 269, 267 
 \bibitem [\protect\citeauthoryear{Gilroy et al.}{1989}]{Gilroy1989} Gilroy K. K., 1989, ApJ, 347, 835
 \bibitem [\protect\citeauthoryear{Girardi et al.}{2000}]{Girardi2000} Girardi L., Bressan A., Bertelli G., Chiosi C., 2000, A\&AS, 141, 371
 \bibitem [\protect\citeauthoryear{Gratton et al.}{1999}]{Gratton1999} Gratton R.G., Carretta, E., Eriksson, K., Gustafsson, B., 1999, A\&A, 350, 955
 \bibitem [\protect\citeauthoryear{Gratton et al.}{2000}]{Gratton2000} Gratton R.G., Sneden C., Carretta E., Bragaglia A. 2000, A\&A, 345, 169
 \bibitem [\protect\citeauthoryear{Grevesse et al.}{2000}]{Grevesse2000} Grevesse N., Sauval A.J., 2000, ``Origin of Elements in the Solar System, Implications of Post-1957 Observations, O. Manuel (ed.), Kluwer, 261
 \bibitem [\protect\citeauthoryear{Gustafsson et al.}{1999}]{Gustafsson1999} Gustafsson B., Karlsson T., Olsson E., Edvardsson B., Ryde N., 1999, A\&A, 342, 426
 \bibitem [\protect\citeauthoryear{Johansson et al.}{2003}]{Johansson2003} Johansson S., Litz\'{e}n U., Lundberg H., Zhang Z., 2003, ApJ, 584, 107
 \bibitem [\protect\citeauthoryear{Lindoff et al.}{1972}]{Lindoff1972} Lindoff U., 1972, A\&AS, 7, 231
 \bibitem [\protect\citeauthoryear{Luck et al.}{1994}]{Luck1994} Luck R. E., 1994, ApJS, 91, 309
 \bibitem [\protect\citeauthoryear{Mashonkina \& Gehren}{2000}]{Mashonkina2000} Mashonkina L. \& Gehren T, 2000, A\&A, 364, 249
 \bibitem [\protect\citeauthoryear{McWilliam}{1998}]{McWilliam1998} McWilliam A. 1998, AJ, 115, 1640
 \bibitem [\protect\citeauthoryear{Meibom}{2000}]{Meibom2000} Meibom S., 2000, A\&A, 361, 929
 \bibitem [\protect\citeauthoryear{Meibom et al.}{2002}]{Meibom2002} Meibom S., Andersen J., Nordstr\"{o}m B., 2002, A\&A, 386, 187
 \bibitem [\protect\citeauthoryear{Mikolaitis et al.}{2010}]{Mikolaitis2010} Mikolaitis \v{S}., Tautvai\v{s}ien\.{e} G., Gratton R., Bragaglia A., Carretta E., 2010, MNRAS, 407, 1866
 \bibitem [\protect\citeauthoryear{Nissen et al.}{1988}]{Nissen1988} Nissen P. E., 1988, A\&A, 199, 146
 \bibitem [\protect\citeauthoryear{Pasquini et al.}{2004}]{Pasquini2004}Pasquini L., Randich S., Zoccali M., Hill V., Charbonnel C., Nordstr\"{o}m B., 2004, A\&A, 424, 951
 \bibitem [\protect\citeauthoryear{Pace et al.}{2008}]{Pace2008} Pace G., Pasquini L., Fran\c{c}ois P., 2008, A\&A, 489, 403 
 \bibitem [\protect\citeauthoryear{Santos et al.}{}]{Santos} Santos N.C., Lovis C., Pace G., Melendez J., Naef D., 2009, A\&A, 493, 309
 \bibitem [\protect\citeauthoryear{Shi et al.}{2002}]{Shi2002} Shi J. R., Zhao G., Chen Y. Q., 2002, A\&A, 381, 982
 \bibitem [\protect\citeauthoryear{Smiljanic et al.}{2009}]{Smiljanic2009} Smiljanic R., Gauderon R., North P., Barbuy, B., Charbonnel C., Mowlavi N., 2009, A\&A, 502, 267
 \bibitem [\protect\citeauthoryear{Stern}{1960}]{Stern1960} Stern, M. E. 1960, Tellus, 12, 172
 \bibitem [\protect\citeauthoryear{Tautvai\v{s}ien\.{e} et al.}{2000}]{Tautvaisiene2000} Tautvai\v{s}ien\.{e} G. Edvardsson B., Tuominen I., Ilyin I., 2000, A\&A, 360, 499
 \bibitem [\protect\citeauthoryear{Tautvai\v{s}ien\.{e} et al.}{2005}]{Tautvaisiene2005} Tautvai\v{s}ien\.{e} G. Edvardsson B., Puzeras E., Ilyin I., 2005, A\&A, 431, 933
 \bibitem [\protect\citeauthoryear{Ulrich}{1972}]{Ulrich1972} Ulrich, R. K. 1972, ApJ, 172, 165
\end{thebibliography}
\end{document}